# Single-Molecule Junction Conductance through Diaminoacenes

Jordan R. Quinn, Frank W. Foss Jr., Latha Venkataraman*, Mark S. Hybertsen, and Ronald Breslow*

*Department of Chemistry, Department of Physics and Department of Applied Physics and Applied Mathematics, and Center for Electron Transport in Molecular Nanostructures, Columbia University, New York, NY 10027, and Center for Functional Nanomaterials, Brookhaven National Laboratory, Upton, NY 11973*

**RECEIVED DATE (automatically inserted by publisher)**; email: lv2117@columbia.edu; rb33@columbia.edu

The study of electron transport through single molecules is essential to the development of molecular electronics.[1] Indeed, trends in electronic conductance through organic nanowires have emerged with the increasing reliability of electron transport measurements at the single-molecule level. Experimental and theoretical work has shown that tunneling distance, HOMO-LUMO gap and molecular conformation influence electron transport in both saturated and π-conjugated nanowires.[2,3] However, there is relatively little experimental data on electron transport through fused aromatic rings.[4] Here we show using diaminoacenes that conductivity depends not only on the number of fused aromatic rings in the molecule, which defines the molecular HOMO-LUMO gap, but also on the position of the amino groups on the rings. Specifically, we find that conductance is highest with minimal disruption of aromaticity in fused aromatic nanowires.

We recently reported on the improved reliability and reproducibility of conductance measurements using amines instead of thiols or isonitriles in metal-molecule-metal junctions.[5] Junctions are formed by breaking Au point contacts in a solution of diamines. Conductance measurements for diaminoacenes were recorded in 1,2,4-trichlorobenzene solution by repeatedly forming and breaking Au point contacts with a modified STM tip[6] (Figure 1b, inset). Typically, diamines were sublimed under vacuum prior to use. Conductance traces measured as a function of tip-sample displacement reveal quantized conductance steps observed at integer multiples of $G_0$ ($2e^2/h$), the fundamental quantum of conductance. Many of the traces reveal steps at molecule-dependent conductance values below $G_0$ (Figure 1a) due to conduction through a single molecule bridging the gap between the two Au point-contacts. Repeated measurements give a statistical assessment of the junction properties presented as conductance histograms (Figure 1b) with the peak representing the most probable measured conductance value for the molecular junction.

Figure 1b shows representative conductance histograms resulting from several thousand conductance traces for three molecules, 1,4-diaminobenzene, 2,6-diaminonaphthalene and 2,6-diaminoanthracene along with a control histogram measured in the solvent alone. Peaks in the conductance histograms determined by Lorentzian fits to the data correspond to the most prevalent single molecule junction conductance. Figure 2 shows the measured conductance values for 5 different acenes. For the upper series, 1,4-diaminobenzene, 1,4-diaminonaphthalene and 9,10-diaminoanthracene, the conductance increases with increasing number of benzo rings, whereas in the lower series, 1,4-diaminobenzene, 2,6-diaminonaphthalene and 2,6-diaminoanthracene, the conductance decreases with increasing number of rings.

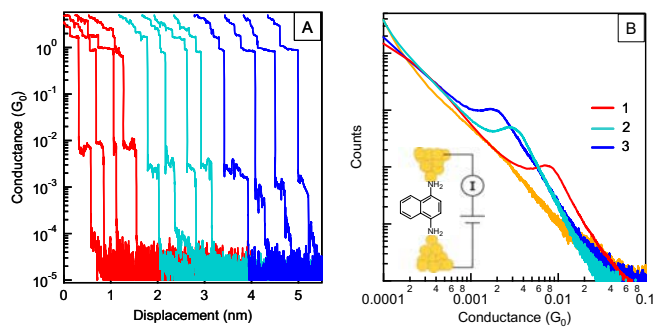

*Figure 1.* Representative conductance traces (a) and histograms (b) for 1,4-diaminobenzene (**1**), 2,6-diaminonaphthalene (**2**) and 2,6-diaminoanthracene (**3**). The lower inset illustrates 1,4-diaminonaphthalene binding in the metal-molecule-metal junction.

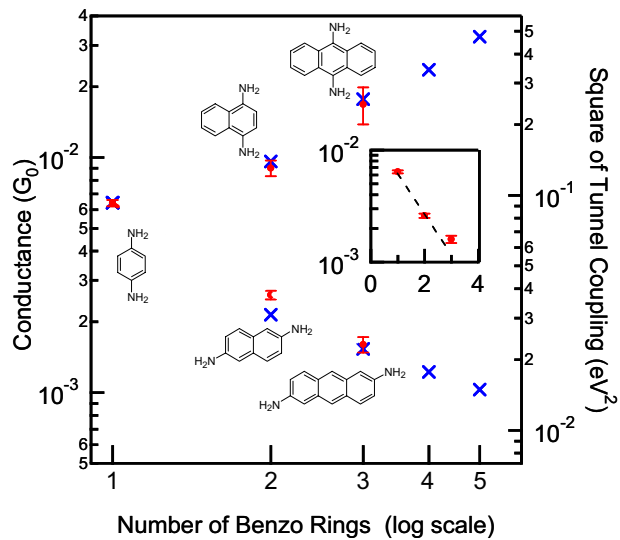

*Figure 2.* Plots of conductance ($G_0$, circles) and square of tunnel coupling (eV$^2$, X's) vs. number of benzo rings (log scale). The inset shows experimental conductance values in the 2,6-diamine series with a linear abscissa.

We have demonstrated previously[3] that the low bias conductance through polyphenyls attached to gold electrodes with amino groups is through a non-resonant tunneling process. Quantitative trends in the junction conductance have been analyzed using quantum chemistry calculations based on density functional theory for amines coupled to Au clusters to represent the electrodes.[7-9] Frontier orbitals on the contact Au atoms are tunnel coupled through the molecular backbone resulting in a symmetric and antisymmetric pair split by 2t. The conductance is proportional to the square of the splitting,[10] well representing trends observed in the conductance through amine linked junctions.[3,5,11] The tunnel coupling was calculated for all the acenes in the experimental study with the results shown in Fig. 2. The trends for both series are accurately represented, strongly suggesting that non-resonant tunneling is responsible for the conductance measured in the acenes.

Interestingly, the 2,6 series conductance does not display the exponential dependence on length normally associated with tunneling.[12] The inset in Fig. 2 shows a clear deviation and the calculations for longer acenes suggest a power law dependence (main panel, roughly $r^{-1}$). Most molecular bridges studied to date (alkanes, oligophenyls, oligophenyleneethynylene, etc.) exhibit modest changes in ionization potential (or HOMO energy) with length and a robust gap to the LUMO energy. In the McConnell picture the donor and acceptor levels (electrode Fermi energies) maintain a roughly constant separation from the bridge energy level independent of bridge length resulting in the exponential conductance. However, the acenes show a rapid reduction in ionization potential with increasing fused ring number. This has significant consequences, as suggested by a simple Hückel model calculation with constant π electron coupling parameter $V$ (see supporting information). If the electrodes are electronically coupled ($\Gamma$) at the 2 and 6 positions and the Fermi energies are aligned to the center of the HOMO-LUMO gap, the conductance does not drop exponentially with length (ring number, r).

$$G = G_0(r+1)^2\Gamma^2V^2/((r+1)^2V^2 + \Gamma^2/4)^2$$

In a similar way, the rise in conductance measured transverse to the acene (e.g. 9,10-diaminoanthracene) indicates systematic approach of the bridge HOMO to the electrode Fermi energy with increasing length. The quantitative trends observed in Fig. 2 depend on the final details of this energy alignment.

The conductivities follow the expected changes in aromatic stabilization when the diamines are oxidized to the dications (Figure 3). With 9,10-diaminoanthracene, there is an increase from one to two benzene rings on such oxidation, while with 2,6-diaminoanthracene all benzenoid rings are lost on such oxidation to the dication.[13] While this is a formalism, and the dications are probably not formed during electrical conduction, we propose that electron abstraction by the drain (anode) runs ahead of the electron donation by the source (cathode) with such electron rich molecular wires. Thus, the wires will be positively charged and have some of the character of the fully oxidized dications. This picture is consistent with the observed substituent dependence of 1,4-diaminobenzene conductance.[11] This idea explains the higher conductivity of 1,4-diaminonaphthalene than of 1,4-diaminobenzene, and the lower conductivity of 2,6-diaminonaphthalene whose dication would have lost the aromaticity of both rings.

To understand further the role of aromatic stabilization and conjugation in electron transport we examined other diaminoacenes. A common feature of the acenes in Figure 2 is that they have amino groups at each end of a string of conjugated double bonds in the Hückel structures shown. This relationship between linking groups is usually associated with higher conductivity.[14] Our measurements on 1,3-diaminobenzene and 2,7-diaminonaphthalene, which lack a string of conjugated double bonds, showed a very small conductance enhancement over a broad range, and no clear conductance peak down to the noise limit of our set-up (see supporting information). These results correspond with our DFT calculations that predict low conductivities for 1,3-diaminobenzene and 2,7-diaminonaphthalene. Interestingly, 1,5-diaminonaphthalene did not display a clear conductance peak either, even though it contains a string of conjugated double bonds between amino groups. The reasons for the lack of a clear conductance peak in 1,5-diaminonaphthalene are being examined further.

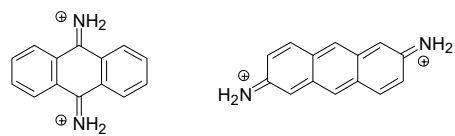

*Figure 3.* Upon oxidation to the dication, 9,10-diaminoanthracene gains a benzene ring (left structure) whereas 2,6-diaminoanthracene loses a benzene ring (right structure).

In addition to establishing length-dependent conductances for diaminoacenes, this work demonstrates the importance of Hückel structures and aromatic stabilization in electron transport. Ongoing investigations are focused on determining the relationship between conductance and electrochemical properties of diaminoacenes. Our preliminary results indicate a relationship between oxidation potential and junction conductance.

**Acknowledgement.** This work was supported primarily by the Nanoscale Science and Engineering Initiative of the National Science Foundation (NSF) under NSF award no. CHE-0117752 and CHE-0641523 and by the New York State Office of Science, Technology, and Academic Research (NYSTAR). This work was supported in part by the US Department of Energy, Office of Basic Energy Sciences, under contract number DE-AC02-98CH10886.

**Supporting Information Available:** Experimental details, conductance histograms and theoretical methods. This material is available free of charge via the Internet at http://pubs.acs.org.